\documentclass[aps,prd,twocolumn,groupedaddress]{revtex4}  

\usepackage{graphicx}
\usepackage{booktabs}
\usepackage{amssymb,bm,mathrsfs,bbm,amscd}
\usepackage[tbtags]{amsmath}

\begin{document}

\preprint{Submitted to prd}

\title{Light propagation in the field of the N-body system and the application in the TianQin mission}

\author{Cheng-Gang Qin}
\author{Yu-Jie Tan}
\author{Ya-Fen Chen}
\author{Cheng-Gang Shao}\email[E-mail:]{cgshao@hust.edu.cn}

\affiliation
{MOE Key Laboratory of Fundamental Physical Quantities
Measurement $\&$ Hubei Key Laboratory of Gravitation and Quantum Physics, PGMF and School of Physics, Huazhong University of Science and Technology,
Wuhan 430074, People's Republic of China}

\date{\today}

\begin{abstract}
Given the high-precision modern space mission, a precise relativistic modeling of observations is required.
By solving the eikonal equation with the post-Newtonian approximation, the light propagation is determined by the iterative method in the gravitational field of an isolated, gravitationally bound N-body system.
Different from the traditional $N$ bodies that are independent with each other in the system, our system includes the velocities, accelerations, gravitational interactions and tidal deformations of the gravitational bodies.
The light delays of these factors then are precisely determined by the analytical solutions.
These delays are significant and are likely to reach a detectable level for the \emph{strong} gravitational fields, such as binary pulsars and some gravitational wave sources.
The result's application in the vicinity of the Earth provides a relativistic framework for modern space missions.
From the relativistic analysis in the TianQin mission,
we find the possible tests for the alternative gravitational theories, such as a possible determination for the post-Newtonian parameter $\gamma$ in the level of some scalar-tensor theories of gravity.
\end{abstract}

\maketitle

\section{Introduction}

In modern times, the growing accuracy of radioscience, laser and astrometric observations requires a detailed modeling of the light propagation in the curved spacetime. By solving the null geodesic equations or eikonal equations in the given metric, the problem of light propagation has been explored within several investigations, e.g., the light propagation in the gravitational field of N arbitrarily moving bodies in first post-Newtonian (1PN) and 1.5PN approximations \cite{pn1,pn2}, the 2PN effects of one arbitrarily moving pointlike body \cite{pn3}, and the post-Minkowskian (PM) effects in the gravitational field of static bodies endowed with arbitrary intrinsic mass-multipole and spin-multipole moments  \cite{pm1,pm2,pm3}.
Furthermore, based on the methods of the Synge world function \cite{sw} and time transfer function (TTF) \cite{ttf1,ttf2}, the light travel time has been studied, such as the 3PM TTF in the field of a static monopole \cite{3pm}.
These solutions are useful when dealing with the light propagation in the Solar-System gravitational field.
When considering the realistic celestial objects, the nonrigid characteristic of bodies cannot be ignored, which could lead to tidal deformations in the N-body system. It means that the gravitational interactions and tidal deformations must be taken into account for the high-precision observables and experiments. The corresponding calculations for the light travel time have not been reported yet. They should be treated carefully since their contributions may be nonnegligible for the space missions or astrometric observations in the foreseeable future.
In particular, the influences due to tidal deformations are much significant in the binary pulsar system and some recent gravitational wave (GW) sources.

The first direct observation of GW events (GW150914) made by the Advanced LIGO opens the era of observational GW astronomy \cite{1,2,thr3} and marks the beginning of a new era in gravitational physics \cite{p1,p2}. Subsequently, several more GW events have been detected by the advanced LIGO and advanced VIRGO \cite{3,4}. At present, the ground-based GW detectors have the performance to detect GWs in the high-frequency regime (10 to $10^{3}$Hz). Aiming to provide more observations of GW events and complement ground-based detectors, the space-based detectors are developed to make GW observations in the low-frequency regime ($10^{-4}$ to 1 Hz) in which they require an accuracy of picometer for the range.

An alternative gravitational wave mission in space may use heliocentric or geocentric orbits for spacecrafts. For the former orbit option, the most representative one is the LISA mission \cite{5}.
For the latter option, early versions SAGITTARIUS \cite{sa} and OMEGA \cite{om} were proposed to ESA in 1993 and to NASA in 1996, respectively. Recently, the TianQin mission was proposed by Luo $et$ $al$ \cite{6}. It relies on an equilateral triangular constellation with inter-spacecraft distance about $1.7 \times 10^{5}$km, which is planned to be launched in the year 2035. Comparing to heliocentric option, these geocentric option missions have several main advantages in the aspects of propulsion requirement, telecommunication system and time requirement for injecting final orbit.
Unfortunately, they are confronted with two technological issues -- the need for keeping sunlight from getting into the telescopes and the need for generating an extremely stable clock frequency. For TianQin, the first issue is cured by a relatively short science run \cite{6}. Another issue comes from the influence of the Earth-Moon-system gravitational field that remains to be solved.

The bigger effects due to the Earth-Moon system are the main difference between heliocentric and geocentric options. The extremely stable clocks used in the GW space mission are significantly affected by the gravitational field and geocentric orbit. The orbit-parameterized frequency shift between spacecrafts should be carried out a full general relativistic treatment in the Solar System barycentric coordinate reference system (BCRS).
Moreover, Time-Delay Interferometry (TDI) \cite{tdi0,tdi1,tdi2,tdi3} needs the accurate knowledge of light propagation delays. And in a simulation code, time delays should be generated as realistically as possible. These require us to rigorously determine the light propagation in the field of Solar System.
In the DSX formalism \cite{dsx1,dsx2}, relativistic phase in the field of N-body system can be derived from solving the eikonal equation \cite{7,eik1}.
Different from classical treatment in the static field, the influences of the N-body-system velocity, acceleration, gravitational interaction and tidal deformation also are determined in this formalism.
These formulas can be applied in the calculations for the Earth-Moon system.

For the LISA mission, the propagation delays and frequency shift in the field of Sun have been studied \cite{8}. Further models for range and frequency measurement in geocentric space missions like  GRACE-Follow-On \cite{gra} and ACES \cite{aces} have been developed to analyse relativistic observables. Considering the different satellite configurations or orbital options, it is necessary to perform a detailed relativistic analysis for TianQin.
Meanwhile considering TianQin running, some effects in frequency shift come from the coordinate effects of the BCRS that depend on the coordinate chart. These effects cancel with each other so that they have no detectable effects. These quantities should be avoided in the simulation code and scientific mission. Finally as the additional productions of TianQin, the analysis of the relativistic effects may provide a formalism to test the fundamental physics, such as testing the post-Newtonian parameter $\gamma$ (it describes the measure of the space-time curvature produced by unit rest mass) \cite{pnp} and local lorentz invariance \cite{fp1,fp2}.

The rest of the paper is organized as follows: In Sec.\ref{II}, we discuss light propagation in the field of Solar System and derive the corresponding relativistic solution for light phase. In the Sec.\ref{IV}, we develop the general relativistic phase model for TianQin. Also, the frequency shift between spacecrafts and coordinate effects are discussed. We give our conclusion in the Sec.\ref{VI}. The method of solving eikonal equation is in Appendix.\ref{appeikonal}. In Appendix.\ref{apptime}, we present the instantaneous coordinate distance and corresponding derivatives. In Appendix.\ref{kepler}, some relationships for keplerian orbit are presented.

\section{relativistic phase of the light in the field of Solar System}\label{II}

\subsection{Space-time reference system}\label{reference}

The Einstein general relativity is a covariant theory in which coordinate charts are merely labels, thus the physical observables should be coordinate-independent quantities. It means that one has a wide freedom to choose the coordinate system in describing the outcome of a particular experiment. In fact, any reference system covering the space-time region of the experiment can be used to describe the results of that experiment. However, some available coordinate systems are associated with a particular celestial body or laboratory have important advantages when describing the observations of precision experiments. By using the harmonic gauge conditions and conservation laws, the relativistic, proper reference frame can be determined. In order to conveniently formulate the coordinate picture of the measurement procedure or offer a simpler mathematical description for the experiments under consideration, one should pick a specific coordinate system to model the observables. For the TianQin mission, the choice is the Solar System BCRS.

The BCRS is a particular implementation of a barycentric reference system in the Solar System with the space-time coordinates ${x^{\mu}}\equiv(ct,\textbf{x})$, which has its origin in the Solar System barycenter (SSB). With the recommendations of IAU resolutions\cite{iau,space1}, the metric of the GCRS may be written in the form only depending on two harmonic potentials
\begin{eqnarray}\label{metricb}
  g_{00}\!\!\!\!&&=1-\frac{2w}{c^{2}}+\frac{2 w^{2}}{c^{4}}+O(c^{-6}),\nonumber\\
  g_{0i}\!\!\!\!&&=\frac{4\delta_{ik}w^{k}}{c^{3}}+O(c^{-5}),\\
  g_{ij}\!\!\!\!&&=-\delta_{ij}-\delta_{ij}\frac{2 w}{c^2}-\frac{3\delta_{ij}}{2}\frac{w^{2}}{c^4}+O(c^{-6}),\nonumber
\end{eqnarray}
where $w$ and $\textbf{w}$ respectively are the scalar and vector harmonic potentials written as
\begin{eqnarray}\label{scalarw}
  w\!\!\!\!&&=\sum_{b}\frac{GM_{b}}{r_{b}}\Big{\{}1+\frac{1}{c^{2}}\Big{(}2v_{b}^{2}-\sum_{c\neq b}\frac{GM_{c}}{r_{cb}}-\frac{1}{2}(\textbf{n}_{b}\cdot\textbf{v}_{b})^{2}\nonumber\\
  &&-\frac{1}{2}(\textbf{r}_{b}\cdot\textbf{a}_{b})\Big{)}\Big{\}}+w_{l}+O(c^{-3}),
\end{eqnarray}
\begin{equation}\label{vectorw}
  \textbf{w}=\sum_{b}\frac{GM_{b}}{r_{b}}\Big{(}\textbf{v}_{b}+\frac{(\textbf{s}_{b}\times\textbf{r}_{b})}{2r^{2}_{b}}\Big{)}+O(c^{-2}),
\end{equation}
where $GM_{b}$ is the gravitational constant of the body $b$, $\textbf{s}_{b}$ is the angular momentum per unit of mass of body $b$, $r_{b}=|\textbf{r}_{b}|=|\textbf{x}-\textbf{x}_{b0}|$ with $\textbf{x}_{b0}$ the barycentric position of body $b$, $\textbf{r}_{bc}=\textbf{x}_{c0}-\textbf{x}_{b0}$ is the distance vector pointing to bodies $c$ from $b$, $\textbf{v}_{b}=d\textbf{x}_{b0}/dt$ and $\textbf{a}_{b}=d\textbf{v}_{b}/dt$ are respectively the barycentric velocity and acceleration of body $b$. Lastly, $w_{l}$ contains the contributions from higher gravitational potential coefficients characterizing the shape of body $b$, and $b$ should represent all the bodies in the Solar System.

\subsection{Relativistic phase model}\label{phase}
In framework of general relativity, solving the null geodesic equation is the standard method to obtain all information of light propagation between two point events \cite{ttf2}. For the calculations of light gravitational delays, the different approaches are also available, such as Synge world function \cite{sw}, time transfer functions \cite{ttf2} and eikonal equation \cite{7}. Here, our choice is based on solving eikonal equation, which is closely connected with the method of Synge world function \cite{sw}.
As the discussion of Neil Ashby $et$ $al$ in Ref.\cite{7}, the problem of light time can be deduced to geometrical optics by using the eikonal equation.
It implies that when we consider the light time between two point events, it is enough to solve the problem by using the eikonal equation.
As a scalar function, the phase $\varphi$ of an electromagnetic wave is invariant under a set of general coordinate transformations,
which satisfies the eikonal equation \cite{gra,7,eikonal1}
\begin{equation}\label{1}
{g^{\mu\nu}}{\partial_{\mu}}{\varphi}\partial_{\nu}{\varphi}=0.
\end{equation}
This equation can be derived from Maxwell equations in which the solution $\varphi$ describes the wave front of an electromagnetic wave propagating in the curved spacetime.
To obtain the solution $\varphi(t,\textbf{x})$, we introduce a covector describing the electromagnetic wavefront in the curved spacetime, $K_{\mu}=\partial_{\mu}\varphi$. For light, it satisfies the equation $g_{\mu\nu}K^{\mu}K^{\nu}=0$ with the vector $K^{\mu}=g^{\mu\nu}\partial_{\nu}\varphi$ tangent to the light ray. Assuming that the phase $\varphi(t,\textbf{x})$ is known, one can straightforwardly study the properties of the light.

To find a solution of the eikonal equation, we expand the phase $\varphi$ with the method of perturbation
\begin{equation}\label{phi1}
  \varphi(t,\textbf{x})=\varphi_{0}+\int{k_{\mu}dx^{\mu}}+\varphi_{\text{GF}}(t,\textbf{x}),
\end{equation}
where $\varphi_{0}$ is a constant, $k_{\mu}=k^{0}(1,\mathrm{\textbf{k}})$ satisfying the relation $\eta^{\mu\nu}k_{\mu}k_{\nu}=0$ is a constant null vector along the direction of propagation of the unperturbed electromagnetic plane wave, and $\varphi_{\text{GF}}$ represents the perturbation due to gravitational field. Since we can define time component $k^{0}=\omega/c$ with $\omega$ the constant angular frequency of the unperturbed electromagnetic wave, the vector $\textbf{k}$ ($|\textbf{k}|=1$) is the unit vector along the propagation of the light giving the wave direction. As a consequence of perturbation method of Eq.(\ref{phi1}), the wave vector $K^{\mu}$ of a light in curved spacetime may be expressed as the form
\begin{equation}\label{phik}
K^{\mu}(t,\textbf{x})=g^{\mu\nu}\partial_{\nu}\varphi=k^{\mu}+k^{\mu}_{\text{GF}}{(t,\textbf{x})},
\end{equation}
where $k^{\mu}_{\text{GF}}{(t,\textbf{x})}$ is the perturbation of the wave vector due to the gravitational field.

To determined $\varphi_{\text{GF}}$, we can use the method of asymptotic perturbation theory \cite{7}, which is given in the Appendix.\ref{appeikonal}.
In the metric given by Eq.(\ref{metricb}), the solution of $\varphi_{\text{GF}}$ are given by Eqs.(\ref{ape6}) and (\ref{ape7}).
For the convenience of calculations and presentations, we separate a particular part of perturbation from $\varphi^{(2)}$ and place it into $\varphi^{(1)}$. This part is the contribution due to $G^{2}$ term in scalar potential $w$.
Therefore, $\varphi_{\text{GF}}$ is rewritten as $\varphi_{\text{GF}}=\varphi^{1}_{\text{GF}}+\varphi^{2}_{\text{GF}}$, where $\varphi^{1}_{\text{GF}}$ is the summation of $\varphi^{(1)}$ and above part from $\varphi^{(2)}$.  We assume that a unique light ray connects two point events $(ct_{A},\textbf{x}_{A})$ and $(ct_{B},\textbf{x}_{B})$ with coordinate time relationship $t_{A}<t_{B}$. From Eqs.(\ref{ape6}) and (\ref{ape7}) , $\varphi^{1}_{\text{GF}}$ is expressed as
\begin{equation}\label{phipgf1}
  \varphi^{1}_{\text{GF}}(t,\textbf{x})=-\frac{R_{AB}k_{0}}{2}\int^{1}_{0}\Big{(}\frac{4w}{c^{2}}-\frac{8\textbf{w}\cdot\textbf{N}_{AB}}{c^{3}} \Big{)}d\lambda,
\end{equation}
where integral is calculated along the straight line of ends $\textbf{x}_{A}$ and $\textbf{x}_{B}$ defined by the parametric equations
\begin{equation}\label{paraeq}
  x^{i}=\lambda(x^{i}_{B}-x^{i}_{A})+x^{i}_{A}, \,\,\,\,\,\,0\leq\lambda\leq1,
\end{equation}
and $\textbf{N}_{AB}=(\textbf{x}_{B}-\textbf{x}_{A})/R_{AB}$ with $R_{AB}=|\textbf{x}_{B}-\textbf{x}_{A}|$.
For the calculations in the vicinity of the celestial bodies, tidal deformations of sources should be taken into account, which means that we should consider potential variations caused by the tidal deformations. By using the body's tidal deformations to correct two harmonic potentials, it is recommended to rewrite Eqs.(\ref{scalarw}) and (\ref{vectorw}) with following forms
\begin{eqnarray}\label{potentials2}
  w(t,\textbf{x})=w_{0}(t,\textbf{x})+w_{tid}(t,\textbf{x})\\
  \textbf{w}(t,\textbf{x})=\textbf{w}_{0}(t,\textbf{x})+\textbf{w}_{tid}(t,\textbf{x}),
\end{eqnarray}
where $w_{0}$ and $\textbf{w}_{0}$ are the previous scaler and vector potentials in Eqs.(\ref{scalarw}) and (\ref{vectorw}), respectively. $w_{tid}$ and $\textbf{w}_{tid}$ represent the tidal-deformation potentials in the  N-body system. Since the vector potential effects are much smaller than the scalar potential's, we neglect the contribution of $\textbf{w}_{tid}$. Generally, $w_{tid}$ can be expressed in the form of Love numbers \cite{gwill}
\begin{equation}\label{tidw}
  w_{tid}(t,\textbf{x})=\sum_{b}(k_{2})_{b}\frac{R_{b}^{5}}{r_{b}^{3}}\sum_{c \neq b}\frac{GM_{c}}{2r_{cb}^{3}}[3(\textbf{n}_{cb}\cdot\textbf{n}_{b})^{2}-1],
\end{equation}
where $(k_{2})_{b}$ is the Love number of body $b$, $R_{b}$ is the equatorial radius of body $b$, $\textbf{n}_{cb}=\textbf{r}_{cb}/r_{cb}$ and $\textbf{n}_{b}=\textbf{r}_{b}/r_{b}$. Since $r^{3}_{b}$ appears in the denominator, it's influence is significant only in the vicinity of gravitational body. Inserting it into Eq.(\ref{phipgf1}) allows us to calculate the deformation effects.

Subsequently, we calculate the mass multipoles $w_{l}$. Using the Blanchet-Damour (B-D) moments, $w_{l}$ may be theoretically determined by the distribution of mass and matter currents. As an example, Earth's contribution to $w_{l}$ is represented here. Usually, it is useful to present these moments of Earth as the parameters evaluated by numerically fitting to various kinds of experimental data, such as satellite motion tracking, geodetic measurements and gravimetry etc. An approximate expansion of Earth multipoles $w_{l,E}$ is the spherical harmonic expansion, which is given by \cite{egm8}
\begin{eqnarray}\label{BDearth}
w_{l,E}(\textbf{x})\!\!\!\!\!&&= \frac{GM_{E}}{r_{E}}\sum^{\infty}_{l=2}\sum^{+l}_{k=0}\Big{(}\frac{R_{E}}{r_{E}}\Big{)}^{l}P_{lk}(\cos \theta)\nonumber\\
\!\!\!\!&& \times (\mathcal{C}^{E}_{lk}\cos(k\phi)+\mathcal{S}^{E}_{lk}\sin(k\phi)),
\end{eqnarray}
where $R_{E}$ is the Earth's equatorial radius, $P_{lk}$ are associated with the Legendre polynomials, $\mathcal{C}^{E}_{lk}$ and $\mathcal{S}^{E}_{lk}$ are spherical harmonic coefficients characterizing Earth, and $\mathcal{C}^{E}_{l0}= - J^{E}_{l}$ is the mass multipole moment of the Earth.

To solve Eq.(\ref{phipgf1}), we use Eqs.(\ref{potentials2})-(\ref{BDearth}) to calculate the perturbed phase. The perturbation $\varphi_{\text{GF}}^{1}$ is further written as
\begin{widetext}
\begin{equation}\label{eswphi}
  \varphi_{\text{GF}}^{1}(t,\textbf{x})= -2 R_{AB}k_{0}\int^{1}_{0}\Big{(}\sum_{b}\frac{GM_{b}}{c^{2}r_{b}}\Big{\{}1+\frac{1}{c^{2}}\Big{(}2v_{b}^{2}-\sum_{c\neq b}\frac{GM_{c}}{r_{cb}}-\frac{1}{2}(\textbf{n}_{b}\cdot\textbf{v}_{b})^{2}
  -\frac{1}{2}(\textbf{r}_{b}\cdot\textbf{a}_{b})\Big{)}\Big{\}}+\frac{w_{l}}{c^{2}}+\frac{w_{tid}}{c^{2}}-\frac{2\textbf{w}\cdot\textbf{N}_{AB}}{c^{3}} \Big{)}d\lambda.
\end{equation}
It may be noted that for the mass multipoles $w_{l}$, $J_{2}$ moment is the main contribution, and the higher moments can be neglected since their influences are much smaller than that of quadrupole term.
After some mathematical manipulations, $\varphi_{\text{GF}}^{1}$ is obtained
\begin{eqnarray}\label{eswphi1}
  \varphi_{\text{GF}}^{1}(t,\textbf{x})&&\!\!\!\!\!\!=
 \sum_{b}\Big{\{}-\frac{2GM_{b}k_{0}}{c^{2}} \Big{(}1-\sum_{c\neq b}\frac{GM_{c}}{c^{2}r_{cb}}\Big{)}
  \ln\frac{r_{bA}+r_{bB}+R_{AB}}{r_{bA}+r_{bB}-R_{AB}}+\varphi_{v}^{b}(t,\textbf{x})+\varphi_{\textbf{r}\cdot\textbf{a}}^{b}(t,\textbf{x})+\varphi^{b}_{tid}(t,\textbf{x})\nonumber\\
  &&\!\!\!\!\!\!- \frac{GM_{b}k_{0}R^{2}_{b}R_{AB}J^{b}_{2}}{ c^{2}(r_{bA}r_{bB}+\textbf{x}_{bA}\cdot\textbf{x}_{bB})}
\Big{[}\frac{1-(\textbf{I}_{b}\cdot\textbf{n}_{bA})^{2}}{r_{bA}}+\frac{1-(\textbf{I}_{b}\cdot\textbf{n}_{bB})^{2}}{r_{bB}}
     -\Big{(}\frac{1}{r_{bA}}+\frac{1}{r_{bB}}\Big{)}\frac{[\textbf{I}_{b}\cdot(\textbf{n}_{bA}+\textbf{n}_{bB})]^{2}}{1+\textbf{n}_{bA}\cdot\textbf{n}_{bB}}\Big{]}\nonumber\\
 &&\!\!\!\!\!\!+\frac{4GM_{b}k_{0}}{c^{3}}   \frac{\textbf{R}_{AB}\cdot[\textbf{s}_{b}\times(\textbf{n}_{bA}+\textbf{n}_{bB})]}{(r_{bA}+r_{bB})^{2}-R_{AB}^{2}} \Big{\}} ,
\end{eqnarray}
where $\textbf{I}_{b}$ is the unit vector along the body $b$ rotation axis, $\varphi_{v}^{b}(t,\textbf{x})$ represents correction from the velocity, $\varphi_{\textbf{r}\cdot\textbf{a}}^{b}(t,\textbf{x})$ is the acceleration contribution to phase, and $\varphi^{b}_{tid}(t,\textbf{x})$ comes from the tidal deformations. This equation describes the gravitational contributions of N-body system in phase including the influences of system's mass, oblateness $J_{2}$, velocity, acceleration, gravitational interactions, rotation and tidal deformations.
A farther calculation obtains $\varphi_{v}^{b}(t,\textbf{x})$
\begin{eqnarray}\label{nrphi}
  \varphi_{v}^{b}(t,\textbf{x})&&\!\!\!\!=\sum_{b}\frac{GM_{b}k_{0}}{c^{2}}R_{AB}
  \Big{\{}
    \Big{[}\frac{4\textbf{N}_{AB}\cdot\textbf{v}_{b}}{cR_{AB}}-\frac{4v^{2}_{b}}{c^{2}R_{AB}}+\frac{(\textbf{N}_{AB}\cdot\textbf{v}_{b})^{2}}{c^{2}R_{AB}}\Big{]}\ln\frac{r_{bA}+r_{bB}+R_{AB}}{r_{bA}+r_{bB}-R_{AB}}\\
  &&\!\!\!\!+ \frac{{\left( {{{\textbf{r}}_{bA}} \cdot {{\textbf{v}}_b}} \right)\left[ {2{r_{bA}}\left( {{{\textbf{r}}_{bB}} \cdot {{\textbf{v}}_b}} \right) + \left( {{r_{bB}} - {r_{bA}}} \right)\left( {{{\textbf{r}}_{bA}} \cdot {{\textbf{v}}_b}} \right)} \right]}}{{c^{2}r_{bA}^2r_{bB}^2\left( {\left( {{{\textbf{n}}_{bA}} \cdot {{\textbf{n}}_{bB}}} \right){+}1} \right)}} - \frac{{r_{bA}^2{{\left( {{{\textbf{N}}_{AB}} \cdot {{\textbf{v}}_b}} \right)}^2}\left[ {\left( {{r_{bB}} - {r_{bA}}} \right){+}2{r_{bB}}\left( {{{\textbf{n}}_{bA}} \cdot {{\textbf{n}}_{bB}}} \right)} \right]}}{{c^{2}r_{bA}^2r_{bB}^2\left( {\left( {{{\textbf{n}}_{bA}} \cdot {{\textbf{n}}_{bB}}} \right){+}1} \right)}}
  \Big{\}},\nonumber
\end{eqnarray}
with $\textbf{N}_{AB}=(\textbf{x}_{B}-\textbf{x}_{A})/R_{AB}$. Clearly, it's value is zero for a static gravitational body. The first term is a direct relativistic correction to the Shapiro term in which the largest correction is proportional to the velocity of gravitational body. The latter two terms are indirect corrections to Shapiro delay.
Integrating the acceleration-dependence term, acceleration perturbation in Eq.(\ref{eswphi1}) is given by
\begin{equation}\label{arphi}
  \varphi_{\textbf{r}\cdot\textbf{a}}^{b}(t,\textbf{x})=\sum_{b}\frac{GM_{b}k_{0}}{c^{4}}
  \Big{\{}
  (r_{bB}-r_{bA})(\textbf{N}_{AB}\cdot\textbf{a}_{b})+[(\textbf{r}_{bA}\cdot\textbf{a}_{b})-(\textbf{N}_{AB}\cdot\textbf{a}_{b})(\textbf{N}_{AB}\cdot\textbf{r}_{bA})]
  \ln\frac{r_{bA}+r_{bB}+R_{AB}}{r_{bA}+r_{bB}-R_{AB}}
  \Big{\}}.
\end{equation}
The acceleration contributions also include direct and indirect corrections to Shapiro delay, which don't exist in case of a static gravitational field. Eqs.(\ref{nrphi}) and (\ref{arphi}) are sufficient to describe motion effects in the light propagation delays for the gravitational field of an isolated, gravitationally bound N-body system.

By introducing the impact vector $\textbf{d}_{b}=\textbf{N}_{AB}\times(\textbf{x}_{bA}\times\textbf{N}_{AB})$ with $d_{b}=|\textbf{d}_{b}|$, the term $\varphi^{b}_{tid}$ is
\begin{eqnarray}\label{tidphi}
  \varphi^{b}_{tid}(t,\textbf{x})=&&\!\!\!\!\!\!-k_{0}(k_{2})_{b}R_{b}^{5}\sum_{c,c\neq b}\frac{GM_{c}}{c^{2}r^{3}_{cb}}
  \Big{\{} \frac{r^{2}_{bA}}{r^{2}_{bB}}\frac{(r^{3}_{bB}-r^{3}_{bA})\mathcal{B}_{c1}+3r^{2}_{bA}R_{AB}\mathcal{B}_{c2}+R_{AB}(R_{AB}^{2}+3\textbf{x}_{bA}\cdot\textbf{R}_{AB})\mathcal{B}_{c3}}{d^{4}_{b}r_{bB}}\nonumber\\
  &&\!\!\!\!\!\!+\frac{1}{d_{b}^{2}}(\frac{\textbf{N}_{AB}\cdot\textbf{x}_{bA}}{r_{bA}}-\frac{\textbf{N}_{AB}\cdot\textbf{x}_{bB}}{r_{bB}})  \Big{\}},
\end{eqnarray}
\end{widetext}
where
\begin{equation}\label{b1}
  \mathcal{B}_{c1}=s^{2}_{A}c^{2}_{cA}c_{A}+2c_{cA}c_{c}(1+c_{A}^{2})+2c_{A}(c^{2}_{cA}+c_{c}^{2}),
\end{equation}
\begin{equation}\label{b2}
  \mathcal{B}_{c2}=(1+c_{A}^{2})(c_{cA}^{2}+2c_{A}c_{cA}c_{c})+2c_{A}^{2}c_{c}^{2},
\end{equation}
\begin{equation}\label{b3}
  \mathcal{B}_{c3}=(1+c_{A}^{2})c_{c}^{2}+2c_{cA}^{2}+4c_{A}c_{cA}c_{c},
\end{equation}
with $c_{A}=\textbf{n}_{Ab}\cdot\textbf{N}_{AB}$, $c_{cA}=\textbf{n}_{bA}\cdot\textbf{n}_{bc}$, $c_{c}=\textbf{N}_{AB}\cdot\textbf{n}_{bc}$ and $s_{A}=\sqrt{1-c^{2}_{A}}$. The tidal effects is in direct proportion to Love number $(k_{2})_{b}$ that is dependent on body $b$, and $(k_{2})_{E}$ is about 0.3 for Earth. The more massive source of tidal force will lead to more obvious deformation on the surface of body, which produces a bigger potential variation. Then, the corresponding light delay arising from this potential variation may reach the nonnegligible level. Recently, GW events reported binary neutron star inspiral. In these systems, tidal deformations are significant due to the strong gravitational interaction, as well as in the binary pulsar systems.

Since $\varphi^{1}_{\text{GF}}$ has been determined by above equations, we consider another term $\varphi^{2}_{\text{GF}}$.  $\varphi^{2}_{\text{GF}}$ can be determined by Eq.(\ref{ape7}).
For the sake of discussion, we rewrite it as two parts: the contribution of the square of Newtonian potential $\varphi^{2}_{\text{GF}}$ and the contribution of coupling terms $\varphi^{2}_{c}$. The velocity- and acceleration-term contributions are higher order than the order of $c^{-4}$, which can be safely ignored.
$\varphi^{2}_{\text{GF}}$ consists of the square terms $G^{2}M^{2}_{b}$, which is \cite{sw}
\begin{equation}\label{phiw2}
  \varphi^{2}_{\text{GF}}=\sum_{b}
  {\frac{G^{2}M^{2}_{b}k_{0}R_{AB}}{c^{4}r_{bA}r_{bB}} }\Big{[}\frac{4}{1+\cos\theta_{b}}-\frac{15\theta_{b}}{4\sin\theta_{b}}
\Big{]},
\end{equation}
with $\cos\theta_{b}=\textbf{n}_{bA}\cdot{\textbf{n}_{bB}}$.
For $\varphi^{2}_{c}$, we consider contributions from the terms in $w^{2}$ likes $(GM_{b}/c^{2}r_{b})(GM_{c}/c^{2}r_{c})$ of two sources, and other contributions are the same magnitude. The analytical solution requires cumbersome calculations. For our calculations in the vicinity of one source, such as body $b$, another distance $r_{c}$ may be expressed as the form of
\begin{equation}\label{rtl}
  \frac{1}{r_{c}}=\frac{1}{r_{bc}}+\frac{\textbf{n}_{bc}\cdot\textbf{r}_{b}}{r^{2}_{bc}}+O(r^{-3}_{bc}).
\end{equation}
It allows us to compute the coupling term effect $\varphi^{2}_{c}$
\begin{eqnarray}\label{ctpe}
  \varphi^{2}_{c}&&\!\!\!\!\!\!=-\frac{GM_{b}k_{0}}{4c^{4}}\sum_{c}\frac{GM_{c}}{r_{bc}}\Big{[}\frac{(\textbf{n}_{bc}\cdot\textbf{N}_{AB})(r_{bB}-r_{bA})}{r_{bc}}\nonumber\\
  &&\!\!\!\!\!\!+\frac{r_{bc}+\textbf{x}_{bA}\cdot\textbf{n}_{bc}-(\textbf{x}_{bA}\cdot\textbf{N}_{AB})(\textbf{n}_{bc}\cdot\textbf{N}_{AB})}{r_{bc}}\nonumber\\
  &&\!\!\!\!\!\!\times\ln\frac{r_{bA}+r_{bB}+R_{AB}}{r_{bA}+r_{bB}-R_{AB}}\Big{]}.
\end{eqnarray}
This equation describes the potential-coupling effect when the light signal passes near body $b$. Its contribution is smaller than that of Eq.(\ref{phiw2}), for the calculation in TianQin, which can be neglected.

Finally, using above method, we consider the light propagation between spacecrafts of TianQin (as the description in Sec.\ref{IV}). At the instant of reception on spacecraft $B$, the signal's phase received from the interferometer on the spacecraft $A$ can be expressed as
\begin{eqnarray}\label{phaseb}
  \varphi (t_{B},\textbf{x}_{B})&&\!\!\!\!\!\!=\varphi (t_{A},\textbf{x}_{A}) + \varphi_{\text{GW}}+\varphi_{\text{noise}}+\frac{2\pi}{c}f_{A}\Big{(}\frac{d\tau_{A}}{dt}\Big{)}_{t_A}\nonumber\\
                                 &&\!\!\!\!\!\! \times[c(t_B-t_A)-\mathcal{R}_{AB}(\textbf{x}_{A},\textbf{x}_{B})],
\end{eqnarray}
where $\textbf{x}_{A}=\textbf{x}_{A}(t_{A})$, $\textbf{x}_{B}=\textbf{x}_{B}(t_{B})$, $f_{A}$ is the proper frequency of transmitter on spacecraft $A$, $\mathcal{R}_{AB}$ is the total geodesic distance without considering gravitational waves, $\varphi_{\text{GW}}$ is the phase contribution of gravitational waves, and $\varphi_{\text{noise}}$ is the noise term containing various kinds of noise, such as laser-frequency
noise, clock noise, optical path-length noise etc. Clearly, the perturbations of gravitational field in phase have been merged into $\mathcal{R}_{AB}$.

\subsection{Estimating the magnitudes for different gravitational terms}\label{sdelay}
From the subSec.\ref{phase}, we can estimate the magnitudes of various terms in phase in the context of the TianQin mission.
TianQin's spacecrafts are placed on the orbit around the Earth so that the influences of Earth-Moon system are significant in the laser signal propagation. A convenient method is to split the gravitational delays into the Earth-Moon-system contribution and external contribution (excluding Earth and Moon).
At the instant of reception on spacecraft $B$, the relativistic phase has been determined by Eq.(\ref{phaseb}).
The term $\mathcal{R}_{AB}$ contains the light propagation delays, which is required in TDI. To study various gravitational contributions in $\mathcal{R}_{AB}$, it is convenient to express the term $\mathcal{R}_{AB}$ as
\begin{equation}\label{mrab}
  \mathcal{R}_{AB}=R_{AB}+\Delta^{1EM}+\Delta^{1ext}+\Delta_{tid}+\Delta^{2}_{GF},
\end{equation}
where last four terms represent the gravitational contributions derived from Eqs.(\ref{eswphi1}) (\ref{tidphi}) and (\ref{phiw2}). With nominal orbital parameters of TainQin, we subsequently use their values to evaluate the order of various gravitational terms.

We start with the second term in Eq.(\ref{mrab}). In order to calculate the magnitude, this term can be expressed as
\begin{equation}\label{1em}
  \Delta^{1EM}=\Delta^{1EM}_{m}+\Delta^{1EM}_{J_{2}}+\Delta^{1EM}_{s}+\Delta^{1EM}_{v}+\Delta^{1EM}_{a}.
\end{equation}
The first term in Eq.(\ref{1em}) is Shapiro term from the contribution of Earth's and Moon's mass monopoles, which is given by
\begin{equation}\label{rm}
  \Delta^{1EM}_{m} \approx \frac{2GM_{E}}{c^{2}}\ln\frac{r_{EA}+r_{EB}+R_{AB}}{r_{EA}+r_{EB}-R_{AB}}
  +\frac{2GM_{M}}{c^{2}}\frac{R_{AB}}{d_{M}}
\end{equation}
with $d_{M}=(r_{MA}+r_{MB})/2$. If we consider the gravitational delays in the PPN formalism, this Shapiro term should be revised to $(1+\gamma)\Delta^{1EM}_{m}/2$, which can be used to test PPN parameter $\gamma$.
In the Solar System, the classical experiment is based on Sun's Shapiro delay to test $\gamma$ \cite{cassini}.
For Earth, the Shapiro term is almost a constant about 2.34cm, which is due to the stable triangular constellation of spacecrafts with respect to Earth.
For Moon, the Shapiro delay term is the level of $5\times10^{-5}$m. However, Moon's Shapiro delay is not a constant delay and varies with $d_{M}$. The variation-part amplitude reaches several micrometers with the frequency about
orbital frequency of spacecraft. Assuming that $\theta_{MA/B}=\textbf{n}_{ME}\cdot\textbf{n}_{EA/B}$, Moon's contribution can be expressed as
\begin{equation}\label{10emm}
  \Delta^{1M}_{m} =
  \frac{2GM_{M}}{c^{2}}\frac{R_{AB}}{r_{EM}}\Big{(} 1+\frac{r_{A}\cos\theta_{MA}+r_{B}\cos\theta_{MB}}{2r_{EM}}\Big{)}.
\end{equation}
For the sake of magnitude estimation, we can assume that spacecraft orbit plane coincides with Moon's. Eq.(\ref{10emm}) can be rewritten as
\begin{equation}\label{1emm}
  \Delta^{1M}_{m} =
  \frac{2GM_{M}}{c^{2}}\frac{R_{AB}}{r_{EM}}\Big{(} 1+\frac{r_{A}}{r_{EM}}\cos {\omega_{ms} t}\Big{)},
\end{equation}
where $\omega_{ms}$ is the angular frequency summation between the spacecraft orbit and Moon orbit (for inverse orbital directions). The constant part is about 49$\mu$m and the amplitude of variation part is about 6$\mu$m.
If we compute this term without that approximation (the spacecraft orbit plane coincides with that of the moon), we would get a slightly bigger delay result, which is bigger than the estimate of Eq.(\ref{1emm}) about 0.5$\mu$m. Eq.(\ref{1emm}) is allowable for estimating the magnitude. In the actual calculations and applications, we use Eq.(\ref{rm}).
To demonstrate the influence of Earth-Moon-system Shapiro delay in the TianQin mission, we adopt Fourier analysis for the rough estimations. From the Fourier analysis of Eq.(\ref{rm}), we find that the effects of Earth-Moon-system Shapiro delay lead to a contribution $3\times10^{-13}$m/Hz$^{1/2}$ at 6mHz smaller than TianQin's position measurement accuracy of level of $1$pm/Hz$^{1/2}$ at 6mHz. In the low-frequency regime ($10^{-4}$-1Hz), all contributions of Shapiro delay are smaller than 1pm/Hz$^{1/2}$. Therefore, the influence of the light delays due to the Earth-Moon-system gravitational field is negligible for the TianQin mission in the GW detections.
Moreover, the Shapiro delay may have the potential to improve the accuracy of post-Newtonian parameter $\gamma$ whose current best value of $\gamma=1+(2.1\pm2.3)\times10^{-5}$ is reported by the Cassini mission \cite{cassini}. From the Moon's contribution in Shapiro delay, it gives a 5.6$\mu$m amplitude at the orbital frequency. If the position measurement accuracy can reach 1pm/Hz$^{1/2}$ at $\mu$Hz, the uncertainty in the PPN parameter $\gamma$ can be tested with the accuracy of $1.8\times10^{-7}$, which approaches the level where some scalar-tensor theories of gravity predict that a deviation from GR might be expected \cite{st1,st2}. A smaller deviation at the level of $10^{-9}$ from GR are predicted by the heuristic string-theory arguments that may be tested by the space missions like BEACON \cite{bc} or LATOR \cite{st2}.

The second term $\Delta^{1EM}_{J_{2}}$ in Eq.(\ref{1em}) is the contribution from the mass's quadrupole moment $J_{2}$. Since the Moon's mass is much smaller than Earth's, it is sufficient to neglect Moon's quadrupole moment.
Considering the equilateral triangular constellation, we can adopt values $\textbf{n}_{EA}\cdot\textbf{n}_{EB}=-1/2$ and $r_{EA}=r_{EB}$ for this term. From the penultimate term of Eq.(\ref{eswphi1}), the delay due to Earth's oblateness can be computed as
\begin{eqnarray}\label{rj2}
  \Delta^{E}_{J_{2}}&&\!\!\!\!\!=1.35\times10^{-7}\text{m}\cdot\Big{[}1-\frac{5}{2}(\textbf{I}_{E}\cdot\textbf{n}_{EA})^{2}-\frac{5}{2}(\textbf{I}_{E}\cdot\textbf{n}_{EB})^{2}\nonumber\\
  &&\!\!\!\!\!-4(\textbf{I}_{E}\cdot\textbf{n}_{EA})(\textbf{I}_{E}\cdot\textbf{n}_{EB})\Big{]}.
\end{eqnarray}
It shows that Earth quadrupole contribution to the delay is large enough to be observed.

The third term in Eq.(\ref{1em}) is the delay contributed from the angular momentum of Earth. The last term in Eq.(\ref{eswphi1}) gives
\begin{eqnarray}\label{rs}
 \Delta^E_{s}&&\!\!\!\!\!=-\frac{4GM_{E}}{c^{3}} \frac{\textbf{R}_{AB}\cdot[\textbf{s}_{E}\times(\textbf{n}_{EA}+\textbf{n}_{EB})]}{(r_{EA}+r_{EB})^{2}-R_{AB}^{2}}\nonumber\\
 &&\!\!\!\!\!=1\times10^{-9}\text{m}\cdot\textbf{N}_{AB}\cdot[\textbf{I}_{E}\times(\textbf{n}_{EA}+\textbf{n}_{EB})].
\end{eqnarray}
The Earth-rotation contribution does not exceed 1nm, and we can amplify or inhibit it through choosing optimized orbit. For the ideal situation, it may be used to give a test of gravitomagnetic effects in light propagation.

Next, we look at the fourth term in Eq.(\ref{1em}), which is due to the velocity effect of gravitational sources. Since the velocity and acceleration corrections are higher effect, it is sufficient to use the relationship $r_{EA}=r_{EB}$. From Eqs.(\ref{eswphi1}) and (\ref{nrphi}), the Earth-velocity-dependent term gives
\begin{widetext}
\begin{equation}\label{rv}
 \Delta^{1E}_{v}=\frac{2GM_{E}}{c^{2}}\Big{\{}
   \Big{(}-\frac{2\textbf{N}_{AB}\cdot\textbf{v}_{E}}{c}+\frac{2v^{2}_{E}}{c^{2}}-\frac{(\textbf{N}_{AB}\cdot\textbf{v}_{E})^{2}}{2c^{2}}\Big{)}
  \ln\frac{r_{EA}+r_{EB}+R_{AB}}{r_{EA}+r_{EB}-R_{AB}}
   -\frac{R_{AB}}{c^{2}r_{EB}}[(\textbf{N}_{AB}\cdot\textbf{v}_{E})^{2}+2(\textbf{n}_{EA}\cdot\textbf{v}_{E})(\textbf{n}_{EB}\cdot\textbf{v}_{E})]\Big{\}}.
\end{equation}
From this equation, the first-order velocity contribution $(v_{E}/c)$ to delay is $-4.7$$\mu$m$\cdot\cos{\omega_{s}}t$, where $\omega_{s}$ is the angular frequency of spacecraft with respect to Earth (in this and subsequent subsections, the initial phase value in cosine/sine functions is set to 0). The second-order velocity contribution $(v_{E}/c)^{2}$ is about $(4.1+0.2\cos{2\omega_{s}}t) \times10^{-10}$m.

\begin{table}[!t]
\caption{\label{tab:1} Parameterized estimates of the light propagation delays between TianQin spacecrafts for various gravitational terms. For the sake of simplicity, we set the constant phase value of all cosine functions in Table I to zero, and use a function $f(x_{1},x_{2})=(5/2)[(x_{1}^{2}+x_{2}^{2}]+4x_{1}x_{2}$ where $x_{1}=\textbf{I}_{E}\cdot\textbf{n}_{EA}$ and $x_{2}=\textbf{I}_{E}\cdot\textbf{n}_{EA}$. For the Moon, we consider the computation with the co-plane assumption between spacecraft's and Moon's orbits. $\mathcal{M}_{b}=GM_{b}c^{-2}/r$ describes the gravitational potential of body $b$.   }
\newcommand{\tabincell}[2]{\begin{tabular}{@{}#1@{}}#2\end{tabular}}
\begin{tabular}{lccc}
\hline
\tabincell{l}{Effect} \,\,\,\,\,\,\,\,\,\,\,\,\,\,\,\,\,\,\,\,\,\,\,\,\,\,\,\,\,\,\,\,\,\,\,\,\,\,\,\, &Equation \,\,\,\,\,\,\,\,\,\, &Contribution Source\,\,\,\,\,\,\,\,\,\,\,\,  &Parameterized value \\
\hline
Earth's monopole mass &$\text{Eq}.(\ref{rm})$
  &$\mathcal{M}_{E}$    &2.34cm\\
Earth's velocity   &$\text{Eq}.(\ref{rv})$
  &$\textbf{v}_{E}$ &$-4.7\mu$m$\cdot\cos{\omega_{s}}t$\\
Earth's acceleration   &$\text{Eq}.(\ref{ra})$
  &$\textbf{a}_{E}$ &$3.8\times10^{-14}$m\\
Interaction with Earth   &$\text{Eq}.(\ref{ra})$
  &                 &$-2.3\times10^{-10}$m\\
Moon's monopole mass   &$\text{Eq}.(\ref{1emm})$
  &$\mathcal{M}_{M}$ &49$\mu$m+6$\mu$m$\cdot\cos\omega_{ms}t$\\
Earth's quadrupole moment $J_{2}$   &$\text{Eq}.(\ref{rj2})$
  &$J_{2}$ &$1.35\times10^{-7}$m$\cdot[1-f(x{1},x_{2})]$\\
Earth's  angular momentum  &$\text{Eq}.(\ref{rs})$
  &$\textbf{s}_{E}$ &$1\text{nm}\cdot\textbf{N}_{AB}\cdot[\textbf{I}_{E}\times(\textbf{n}_{EA}+\textbf{n}_{EB})]$\\
Sun's monopole mass   &$\text{Eq}.(\ref{rsj})$
  &$\mathcal{M}_{S}$  &3.3m$\cdot(1+\zeta\cos\omega_{s}t+e_{E}\cos\omega_{E}t)$\\
Deformation due to Moon  &$\text{Eq}.(\ref{rtid})$
  &          &$2.1\text{pm}(12\mathcal{B}_{M2}+30\mathcal{B}_{M3}-1)$\\
Deformation due to Sun  &$\text{Eq}.(\ref{rtid})$
  &          &$1\text{pm}(12\mathcal{B}_{S2}+30\mathcal{B}_{S3}-1)$\\
Sun's 2PN   &
  &                  &$28\text{nm}\cdot(1+2e_{E}\cos\omega_{E}t)$\\
Jupiter's mass   &
  &$\mathcal{M}_{J}$  &$(5-8)\times10^{-4}\text{m}$\\
\hline
\end{tabular}
\end{table}

The last term in Eq.(\ref{1em}) is the acceleration correction to light delay, which contains the contributions from Earth's acceleration and interaction with other bodies. From Eqs.(\ref{eswphi1}) and (\ref{arphi}), it is expressed as
\begin{equation}\label{ra}
   \Delta^{1E}_{a}=\frac{GM_{E}}{c^{4}}
  \Big{[}
  (\textbf{N}_{AB}\cdot\textbf{a}_{E})(\textbf{N}_{AB}\cdot\textbf{r}_{EA})-(\textbf{r}_{EA}\cdot\textbf{a}_{E})-\sum_{c\neq E}\frac{2GM_{c}}{r^{2}_{cE}}r_{cE}\Big{]}
  \ln\frac{r_{EA}+r_{EB}+R_{AB}}{r_{EA}+r_{EB}-R_{AB}}.
\end{equation}
\end{widetext}
Through rough estimations, the acceleration contribution to delay is about $3.8$$\times10^{-14}$m a negligible level, and the delay due to interaction is bigger reaching $-2.3$$\times10^{-10}$ m.

The third term in Eq.(\ref{mrab}) represents external gravitational contribution to the delay, which is mainly due to Sun and other planets. Since these gravitational sources are remote enough, $\Delta^{1ext}$ can be expressed as
\begin{equation}\label{rsj}
  \Delta^{1ext} =
  \sum_{b\neq E,M}
  \Big{(}1-\frac{2\textbf{N}_{AB}\cdot\textbf{v}_{b}}{c} \Big{)}
  \frac{2GM_{b}}{c^{2}}\frac{R_{AB}}{d_{b}}
\end{equation}
with $d_{b}=(r_{bA}+r_{bB})/2$. Assuming that $\omega_{E}$ and $e_{E}$ are respectively the angular frequency and eccentricity of Earth orbit, a direct estimate gives that the Sun's contribution is
3.3m$\cdot(1+\zeta\cos\omega_{s}t+e_{E}\cos\omega_{E}t)$ whereas for Jupiter it is $0.5-0.8$mm. $\zeta=r_{EA}/(1 AU)$ is a constant.

The delay contribution of tidal deformations is given by the fourth term in Eq.(\ref{mrab}). Only taking Earth into account, we obtain
\begin{equation}\label{rtid}
  \Delta_{tid}=\sum_{b \neq E}\frac{\sqrt{3}(k_{2})_{E}GM_{b}R_{E}^{5}}{c^{2}r_{Eb}^{3}d_{E}^{2}}(12\mathcal{B}_{b2}+30\mathcal{B}_{b3}-1),
\end{equation}
where $(k_{2})_{E}$ is Earth's Love number, and $d_{E}$ is the impact parameter with the value about $r_{EA}/2$ or $r_{EB}/2$. Moon and Sun give rise to most of deformations of Earth. The calculation implies that their contribution reaches the level of the picometer and the contribution of Moon's tidal force is about two times contribution of Sun's.
This tidal-deformation delay is more significant when the impact parameter of light with respect to Earth is smaller. If we consider BEACON conception, tidal-deformation delay even can reach several nanometers, which suggests that BEACON could yield a test of this delay with the accuracy of level of $10\%$.

Finally, we evaluate the last term in Eq.(\ref{mrab}), which is the 2PN contribution of Sun mass. Eq.(\ref{phiw2}) yields Sun's 2PN delay
$2.8\times10^{-8}\text{m}\cdot(1+2e_{E}\cos\omega_{E}t)$. Clearly, Earth's contribution in this term is the level of 0.1pm and other bodies contributions are much smaller. Therefore, we just keep Sun's contribution in this term. To summarize, we give a list of the various gravitational delays in the TABLE.\ref{tab:1}.

\section{General relativistic phase model and frequency shift for TianQin}\label{IV}

\begin{figure}
  \centering
  \includegraphics[width=0.48\textwidth]{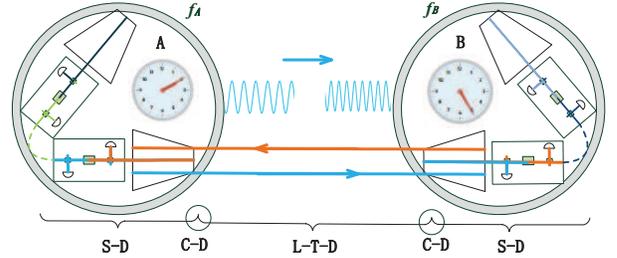}
  \caption{ Schematic diagram of the TianQin laser ranging interferometric measurement. A light signal with frequency $f_{A}$ sent from spacecraft $A$ is received by spacecraft $B$.  The local oscillator or clock on spacecraft compares its frequency $f_{B}$ with received signal frequency that forms frequency observable. Clearly, the frequency shift can be split into spacecraft-dependence (S-D) part, light-trajectory-dependence
  (L-T-D) part, as well as their coupling (C-D) part.  }\label{figure0}
\end{figure}

As the advice of Arthur Schawlow, never measure anything but frequency \cite{nob1,nob2}. Essentially, the measurable quantity of LRI is the frequency difference usually expressed in the form of frequency shift. For the TianQin mission, the frequency shift between spacecrafts $A$ and $B$ linked by laser link may be split into three parts, the light-trajectory-dependence part, spacecraft-dependence (clock-dependence) part and coupling part of light trajectory and spacecraft (FIG.\ref{figure0}). The light-trajectory-dependence part includes the first-order Doppler effect, propagation delay effects, GW signal and higher effects in light trajectory. And spacecraft-dependence part concerns gravitational redshift, gravitomagnetic clock effect, clock noise etc. For the coupling part, the biggest term with the order of $c^{-3}$ comes from the coupling between the gravitational redshift and first-order Doppler effect, and other terms are higher than $c^{-3}$.
From the theoretical point of view, the one-way frequency shift between spacecrafts is characterized as
\begin{eqnarray}\label{piv2}
  &&\!\!\!\!\!\!\frac{\Delta f}{f}(\textbf{v}_{A},t_{A},\textbf{x}_{A},t_{B},\textbf{x}_{B})\\
  &&\!\!\!\!\!\!=\left(\frac{\Delta f}{f}\right)_{GW}+\left(\frac{\Delta f}{f}\right)_{s}+\left(\frac{\Delta f}{f}\right)_{GF}+\left(\frac{\Delta f}{f}\right)_{n}+...\nonumber
\end{eqnarray}
where it is dependent on the velocities and positions of spacecrafts (for more details, gravitational constants of Earth and Moon, angular momentum and other parameters are needed). On the right-hand side of equation, the first term is the gravitational wave effect in frequency shift, the target effect of TianQin. The second term represents the special-relativistic Doppler effect including the first-order and higher-order Doppler effects, and the third term contains all contributions of gravitational field, such as gravitational redshift. Subsequent term describes the noise term mentioned in Eq.(\ref{phaseb}). For the laser frequency noise, we can directly introduce it into noise term, and for the shot noise or phase noise, using Eq.(\ref{phaseb}) may transfer them into this form. The ellipsis includes other possible observable effects in TianQin, such as, possible violation effects of Local Lorentz Invariance and  Local Position Invariance, which will be studied in our further works.

\begin{figure}
  \centering
  \includegraphics[width=0.48\textwidth]{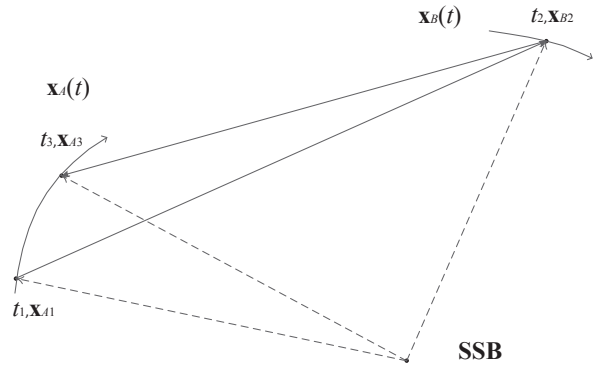}
  \caption{ Schematic diagram of the timing events on TianQin for the signal propagation. $\textbf{x}_{A}(t)$ and $\textbf{x}_{B}(t)$ are respectively the trajectories of spacecrafts $A$ and $B$ almost coincide with each other. The transmission of signal is the event point ($t_{1},\textbf{x}_{A1}$), and its reception is $\textbf{x}_{B2}$ by spacecraft $B$ at time $t_{2}$.
  For the two-way measurement, the signal is to return to spacecraft $A$ at $\textbf{x}_{A3}$.}\label{figure1}
\end{figure}

In order to model the laser ranging interferometric (LRI) observables of TianQin in detail, we consider that the spacecrafts $A$ and $B$ move on their worldlines $\textbf{x}_{A}(t)$ and $\textbf{x}_{B}(t)$, respectively. At coordinate time $t_{1}$, a laser signal with phase $\varphi(t_{1},\textbf{x}_{A1})$ is transmitted by the onboard oscillator of spacecraft $A$, where we set that $\textbf{x}_{Ai}$ represents $\textbf{x}_{A}(t_{i})$, as well as the corresponding quantities $\textbf{x}_{Bi}$ in the following text. At coordinate time $t_{2}$, this signal is received at the spacecraft $B$ ($t_{2},\textbf{x}_{B2}$) with phase $\varphi(t_{2},\textbf{x}_{B2})=\varphi(t_{1},\textbf{x}_{A1})$. The interferometer onboard the spacecraft $B$ compares the phase of local laser oscillator at $t_{2}$ to the phase of the received signal at $\textbf{x}_{B2}$ from the spacecraft $A$. This comparison procedure produces the phase difference and frequency observables from which range and range rate between the two spacecrafts are deduced. These phase and frequency data constitute the GW signal, gravitational field effects, Doppler effects etc.
For the two-way measurement (as shown in FIG.\ref{figure1}), that signal is coherently retransmitted at the spacecraft $B$ ($t_{2},\textbf{x}_{B2}$) with phase $\varphi(t_{2},\textbf{x}_{B2})$ and subsequently is received at the spacecraft $A$ ($t_{3},\textbf{x}_{A3}$) with phase $\varphi(t_{3},\textbf{x}_{A3})=\varphi(t_{2},\textbf{x}_{B2})$. Similarly, the interferometer onboard the spacecraft $A$ compares the phase of local laser oscillator at $t_{3}$ to the phase of the received signal at $\textbf{x}_{A3}$ from the transponder on the spacecraft $B$.

We start our discussion with the one-way measurement. At the spacecraft $B$, the detectable quantity is the difference between the instantaneous local phase and received phase. For one-way measurement, an oscillator onboard spacecraft $A$ with proper frequency $f_{A}$ generates a signal with frequency $f_{A}(\tau_{A1})$ at proper time $\tau_{A1}$. This signal is received by spacecraft $B$ at proper time $\tau_{B2}$ in which the local oscillator's proper frequency is $f_{B}(\tau_{B2})$ at that instant. Then, the infinitesimal difference $\delta\varphi_{AB}(\tau_{B2})$ between the received phase $d\varphi_{AB}(\tau_{B2})$ and the locally generated phase $d\varphi_{B}(\tau_{B2})$ may be expressed by taking difference between two phase values as
\begin{eqnarray}\label{dphiab1}
  \delta\varphi_{AB}(\tau_{B2})&&\!\!\!\!\!\!=d\varphi_{B}(\tau_{B2})-d\varphi_{AB}(\tau_{B2})\nonumber\\
                              &&\!\!\!\!\!\!=2\pi(f_{B}(\tau_{B2})-f_{AB}(\tau_{B2}))d\tau_{B2},
\end{eqnarray}
where $f_{AB}(\tau_{B2})$ is the frequency of the oscillator at spacecraft $A$ measured at spacecraft $B$. The received phase $d\varphi_{AB}(\tau_{B2})$ is originally generated from spacecraft $A$ at proper times $\tau_{A1}$, which may be expressed by proper frequency $f_{A}(\tau_{A1})$ and infinitesimal proper time interval $d\tau_{A1}$ at spacecraft $A$, $d\varphi_{AB}(\tau_{B2})=d\varphi_{A}(\tau_{A1})=
2\pi f_{A}(\tau_{A1})d\tau_{A1}$. The relationship allows us to express the frequency $f_{AB}(\tau_{B2})$ in the form of the frequency $f_{A}(\tau_{A1})$ at the proper time $\tau_{A1}$ on spacecraft $A$
\begin{equation}\label{fab1}
  f_{AB}(\tau_{B2})=\frac{d\tau_{A1}}{d\tau_{B2}}f_{A}(\tau_{A1}).
\end{equation}
Using Eq.(\ref{fab1}), the infinitesimal difference $\delta\varphi_{AB}(\tau_{2})$ of phase can be rewritten in the form of proper frequencies generated locally on spacecrafts $A$ and $B$
\begin{equation}\label{dphiab2}
  \delta\varphi_{AB}(\tau_{B2})=2\pi\Big{(}f_{B}(\tau_{B2})-f_{A}(\tau_{A1})\frac{d\tau_{A1}}{d\tau_{B2}}\Big{)}d\tau_{B2}.
\end{equation}

In fact, spacecraft $B$ will use a phase-locked detection scheme to held confirm the presence of the detection signal with high sensitivity, in which a frequency offset $f_{Bo}(\tau_{B2})$ is brought into locally generated signal. This frequency offset can respond to the received signal for its subsequent retransmission process. The coherent retransmission implies that frequency offset satisfies the relationship
\begin{equation}\label{freoff}
  f_{B}(\tau_{B2})+f_{Bo}(\tau_{B2})=f_{A}(\tau_{A1})\frac{d\tau_{A1}}{d\tau_{B2}}.
\end{equation}
The retransmitted signal is received at proper time $\tau_{A3}$ and compared with locally generated signal on spacecraft $A$. Thus, similar to expression of $\delta\varphi_{AB}(\tau_{2})$, the infinitesimal difference $\delta\varphi_{BA}(\tau_{A3})$ in phase measured on spacecraft $A$ is given by
\begin{eqnarray}\label{dphiba3}
  \delta\varphi_{BA}(\tau_{A3})&&\!\!\!\!\!\!=2\pi\Big{(}  f_{A}(\tau_{A3})  \nonumber\\
                               &&\!\!\!\!\!\!-(f_{B}(\tau_{B2})+f_{Bo}(\tau_{B2}))\frac{d\tau_{B2}}{d\tau_{A3}} \Big{)} d\tau_{A3}.
\end{eqnarray}
Eqs.(\ref{dphiab2}) and (\ref{dphiba3}) can be used to deduce the observational equations, which is needed for processing the scientific data.

For more practical consideration, the LRI observables of TianQin are continuous signal. The continuous changes in phase difference generate time series data. To obtain the changes of phase difference, the proper times should be treated as continuous variable allowing to formally integrate Eqs.(\ref{dphiab2}) and (\ref{dphiba3})
as follows:
\begin{eqnarray}\label{piv1}
  &&\Delta\varphi_{AB}(\tau_{B2})=\int\delta\varphi_{AB}(\tau_{B2}),\\
 && \Delta\varphi_{BA}(\tau_{A3})=\int \delta\varphi_{BA}(\tau_{A3}).
\end{eqnarray}
The two quantities are respectively the LRI observables on the spacecrafts $B$ and $A$ by comparing the phase of local oscillator with the phase of received signal.

In order to develop Eq.(\ref{dphiab2}) or Eq.(\ref{dphiba3}), we should establish the differential equation between the spacecraft proper times ($\tau_{A}$ and $\tau_{B}$ ) and coordinate time in BCRS, which is given by
\begin{equation}\label{taudt}
  \frac{d\tau_{A/B}}{dt}=1-\frac{1}{c^{2}}\Big{[}\frac{\textbf{v}_{A/B}^{2}}{2}+ w(\textbf{x}_{A/B}) \Big{]}+O(c^{-4}).
\end{equation}
We analyse $\delta\varphi_{AB}(\tau_{B2})$ in which the same process can be used for $\delta\varphi_{BA}(\tau_{A3})$. Using Eq.(\ref{taudt}), Eq.(\ref{dphiab2}) is rewritten as following
\begin{eqnarray}\label{rdphiab3}
   \delta\varphi_{AB}(\tau_{B2})&&\!\!\!\!\!\!=2\pi\Big{(}f_{B}(\tau_{B2}) -\\
                               &&\!\!\!\!\!\!f_{A}(\tau_{A1})
                               {\left( {\frac{{d{\tau _A}}}{{dt}}} \right)_{{t_1}}}\left( {\frac{{d{\tau _B}}}{{dt}}} \right)_{{t_2}}^{-1}\frac{{d{t_1}}}{{d{t_2}}}
                               \Big{)}d\tau_{B2}.\nonumber
\end{eqnarray}
Considering the phase property of light propagation, the ratio of coordinate times in this equation may be expressed as
\begin{equation}\label{dt12}
  \frac{dt_{1}}{dt_{2}}=1-\frac{1}{c}\frac{d}{dt_{2}} \mathcal{R}_{AB}(\textbf{x}_{A1},\textbf{x}_{B2}).
\end{equation}
By introducing the instantaneous coordinate distance $D_{AB}=|\textbf{x}_{B2}-\textbf{x}_{A2}|$ (Appendix.\ref{apptime}), this equation is further written as
\begin{eqnarray}\label{dtr12}
   \frac{dt_{1}}{dt_{2}} &&\!\!\!\!\!\! =  1-\frac{1}{c}\frac{d}{dt_{2}}\Big{\{} {D_{AB}} + \Delta^{GF}_{AB}+\frac{{{{\bf{D}}_{AB}} \cdot {{\bf{v}}_A}}}{c}\nonumber\\
          &&\!\!\!\!\!\! + \frac{{{D_{AB}}}}{{2{c^2}}}\Big{[} {{{{\bf{v}}_A^2 -\bf{D}}_{AB}} \cdot {{\bf{a}}_A} + \frac{{{{({{\bf{D}}_{AB}} \cdot {{\bf{v}}_A})}^2}}}{{D_{AB}^2}}} \Big{]} \Big{\}}.
\end{eqnarray}
Apparently, the light-trajectory-dependence part in frequency shift may be described by this equation, such as, the first-order Doppler effect and Sagnac effect.
The second term in the parenthesis describes the frequency shift due to the gravitational delay. Using the Shapiro delay term gives
\begin{eqnarray}\label{dtiv8}
  \frac{d \Delta^{GF}_{AB}}{cdt}&&\!\!\!\!\!\!=\sum_{b}
  \frac{2GM_{b}}{c^{3}r_{bA}r_{bB}}\Big{[}\frac{(r_{bA}+r_{bB})\textbf{N}_{AB}\cdot\textbf{v}_{AB}}{1+\textbf{n}_{bA}\cdot\textbf{n}_{bB}}\nonumber\\
  &&\!\!\!\!\!\!-\frac{(\textbf{n}_{bA}\cdot\textbf{v}_{bA}+\textbf{n}_{bB}\cdot\textbf{v}_{bB})R_{AB}}{1+\textbf{n}_{bA}\cdot\textbf{n}_{bB}}\Big{]}.
\end{eqnarray}
The contribution of Earth's gravitational delays is about $4\times10^{-15}e$, where $e$ is the eccentricity of spacecraft with respect to Earth.

Inserting Eq.(\ref{taudt}) into Eq.(\ref{rdphiab3}), we have
\begin{eqnarray}\label{iv9}
  &&{\left( {\frac{{d{\tau _A}}}{{dt}}} \right)_{{t_1}}}\left( {\frac{{d{\tau _B}}}{{dt}}} \right)_{{t_2}}^{-1}\nonumber\\
  &&=1+\frac{1}{c^{2}}
  \left( \frac{\textbf{v}_{B}^{2}-\textbf{v}_{A}^{2}}{2} + w_{B}-w_{A}     \right) +O(c^{-4}).
\end{eqnarray}
It constitutes the most of clock-dependence part in frequency shift.
Estimating this equation, the second-order Doppler effect is about $1\times10^{-9}$ and the gravitational redshift due to Sun is about $1\times10^{-11}$.  However, observable physical quantities are much smaller.

Considering TianQin's constellation, the second-order Doppler term can be reexpressed as
\begin{eqnarray}\label{iv10}
 \frac{1}{c^{2}} \left( \frac{\textbf{v}_{B}^{2}-\textbf{v}_{A}^{2}}{2}\right )
  &&\!\!\!\!\!=\frac{1}{c^{2}} \frac{d}{dt}(\textbf{R}_{AB}\cdot\textbf{v}_{E})-\frac{\textbf{R}_{AB}\cdot\textbf{a}_{E}}{c^{2}}       \nonumber\\
 &&\!\!\!\!\! +\frac{1}{2c^{2}}(\textbf{v}^{2}_{EB}-\textbf{v}^{2}_{EA}).
\end{eqnarray}
The first term essentially is the coordinate effect that will cancel with the third term in the parenthesis of Eq.(\ref{dtr12}). The cancellation leads to measurable effects in that term of Eq.(\ref{dtr12}) only dependent the spacecraft's velocity with respect to Earth. The magnitude of the second term is $-1\times10^{-11}$, which has equivalent value but opposite sign with gravitational refdshift.
Further, the gravitational redshift term can be rewritten as
\begin{eqnarray}\label{iv11}
  \frac{w_{B}-w_{A}}{c^{2}} &&\!\!\!\!\!=
  \frac{{\sqrt 3 G{M_E}e}}{{{c^2}a}}\left(  1+\frac{{3R_{E}^2}}{{2{a^2}}}{J_2} \right) \cos {\omega _s}t\nonumber\\
   &&\!\!\!\!\!+\frac{1}{c^{2}}\sum_{b\neq E}\frac{GM_{b}}{r_{bE}^{3}}\textbf{x}_{bE}\cdot\textbf{R}_{BA}+O(e^{2},r_{bE}^{-3}),
\end{eqnarray}
with semimajor axis $a=10^{5}$km. Clearly, the first term is the Earth's gravitational redshift, in which the contribution of mass monopole is about $7.7e\times10^{-11}$
whereas for quadrupole moment it is $4.7e\times10^{-16}$.
The second term represents gravitational redshift due to other body's gravitational field, which would cancel with the second term in Eq.(\ref{iv10})
since the Earth's acceleration is given by $\nabla\sum_{b\neq E} (GM_{b})/r_{bE}$. Thus, the influence of the Sun and Moon gravitational field only is given in the form of tidal potential. This is the characteristic of geocentric orbit option. The influences of other bodies (except for Earth) become
\begin{equation}\label{iv12}
  \frac{u^{t}_{B}-u^{t}_{A}}{c^{2}}\simeq\sum_{b\neq E}\frac{3GM_{b}a^{2}}{2c^{2}r^{3}_{bE}}[(\textbf{n}_{bE}\cdot\textbf{n}_{EB})^{2}-(\textbf{n}_{bE}\cdot\textbf{n}_{EA})^{2}],
\end{equation}
where $u^{t}$ is defined by tidal potential. This equation is the ignored term in Eq.(\ref{iv11}) involving $r_{bE}^{-3}$. The contributions due to Moon and Sun are $1\times10^{-14}$ and $6\times10^{-15}$, respectively. The other body contributions are even smaller and can be omitted (e.g. it is the level of $10^{-20}$ for Jupiter).

Subsequently, Earth's tidal deformation is taken into account. From Eq.(\ref{tidw}), its effect in frequency shift $(\delta f/f)_{tid}$ is obtained, which is given in the form of Love number
\begin{equation}\label{frtid}
  \left(\frac{\delta f}{f} \right)_{tid}=\sum_{c \neq E}\frac{3(k_{2})_{E}GM_{c}R_{E}^{5}}{2c^{2}r_{cE}^{3}a^{3}}[(\textbf{n}_{cE}\cdot\textbf{n}_{EB})^{2}-(\textbf{n}_{cE}\cdot\textbf{n}_{EA})^{2}].
\end{equation}
The tidal force sources of deformations mainly are Sun and Moon that lead to frequency shift with a negligible level of $10^{-20}$. This effect grows with a lower orbit. The estimate implies that the frequency shift due to tidal deformations may reach a measurable level in the binary pulsar systems.

For the coupling terms, we consider frequency shift due to the coupling between Earth's gravitational redshift and Doppler effect, which is about $5\times10^{-17}e^{2}$.  Whereas for coupling term involving Sun it is smaller. The coupling effects for TianQin are negligible.

\begin{table}[!t]
\caption{\label{tab:2} Parameterized estimates of the one-way frequency shift between spacecrafts for the TianQin mission. We have set the original phase in sine and cosine functions to 0. And the angles are given by
$\cos\theta_{bA/B}=\textbf{n}_{bE}\cdot\textbf{n}_{EA/B}$, where $b$ represents Moon or Sun.   }
\newcommand{\tabincell}[2]{\begin{tabular}{@{}#1@{}}#2\end{tabular}}
\begin{tabular}{lccc}
\hline
\tabincell{l}{Effect} &Parameterized value  \\
\hline
First-order Doppler effect  &$5.8\times10^{-6}\cdot e\sin\omega_{s}t$ \\
Second-order Doppler  effect  &$7.7\times10^{-11}\cdot e\cos {\omega _s}t$ \\
Earth's mass monopole  &$7.7\times10^{-11}\cdot e\cos {\omega _s}t$ \\
Earth's mass $J_{2}$ term   &$4.7\times10^{-16}\cdot e\cos {\omega _s}t$   \\
Moon's mass monopole            &$1\times10^{-14}(\cos^{2}\theta_{MB}-\cos^{2}\theta_{MA})$\\
Sun's mass monopole &$6\times10^{-15}(\cos^{2}\theta_{SB}-\cos^{2}\theta_{SA})$\\
\hline
\end{tabular}
\end{table}

After above mathematics manipulations and using Eq.(\ref{app5}), the infinitesimal difference $\delta\varphi_{AB}$ becomes
\begin{widetext}
\begin{eqnarray}\label{fafb}
\delta\varphi_{AB}(\tau_{B2})&&\!\!\!\!\!\!=2\pi\Big{\{}f_{B}(\tau_{B2}) - f_{A}(\tau_{A1})+f_{A}(\tau_{A1})\Big{[}\frac{\textbf{N}_{AB}\cdot\textbf{v}_{AB}}{c}- \frac{\textbf{v}_{EB}^{2}-\textbf{v}_{EA}^{2}}{2c^{2}}\nonumber\\
&&\!\!\!\!\!\!+\frac{1}{c^{2}}((\textbf{N}_{AB}\cdot\textbf{v}_{EA})(\textbf{N}_{AB}\cdot\textbf{v}_{EB})-(\textbf{N}_{AB}\cdot\textbf{v}_{EA})^{2}
  +\textbf{D}_{AB}\cdot\textbf{a}_{EA})\nonumber\\
  &&\!\!\!\!\!\!-  \frac{{\sqrt 3 G{M_E}e}}{{{c^2}a}}\left(  1+\frac{{3R_{E}^2}}{{2{a^2}}}{J_2} \right) \cos {\omega _s}t
  -\sum_{b\neq E}\frac{3GM_{b}a^{2}}{2c^{2}r^{3}_{bE}}[(\textbf{n}_{bE}\cdot\textbf{n}_{EB})^{2}-(\textbf{n}_{bE}\cdot\textbf{n}_{EA})^{2}] \Big{]}\Big{\}}d\tau_{B2}.
\end{eqnarray}
\end{widetext}
The $c^{-1}$ term is the first-order Doppler effect, which may be written in the  orbit-parameter form
\begin{equation}\label{iv13}
 - \frac{\textbf{N}_{AB}\cdot\textbf{v}_{AB}}{c}=\frac{\sqrt{3}e}{2c}\sqrt{\frac{GM_{E}}{a}}\sin\omega_{s}t=5.8\times10^{-6}\cdot e\sin\omega_{s}t.
\end{equation}
For the Kepler orbit, the velocity is sufficient to the relation $\textbf{v}^{2}=2U+K$, where $U$ is Newtonian gravitational potential and $K$ is a constant. The clock's second-order Doppler effect can be written as
\begin{equation}\label{iv14}
   \frac{\textbf{v}_{EB}^{2}-\textbf{v}_{EA}^{2}}{2c^{2}}=\frac{{\sqrt 3 G{M_E}e}}{{{c^2}a}} \cos {\omega _s}t.
\end{equation}
This term has the same magnitude with the Earth's gravitational redshift, which is a characteristic of Keplerian orbit. In fact, the practical orbit deviates from Keplerian orbit, which also causes relativistic effects between spacecraft's clocks. These effects may need to be considered for a long-time integrate that can be estimated by perturbed Kepler orbit \cite{cqg2019}. After the time interval of a complete orbit period (3.65 days), the relative frequency difference between spacecraft's clocks may reach the level of $10^{-13}e$. To summarize, several important effects are listed in TABLE.\ref{tab:2}.

\section{Conclusion}\label{VI}

The high-precision space missions ask for modeling the precise relativistic observations of the laser and frequency measurements.
By solving eikonal equation in the BCRS with the post-Newtonian approximation, the light propagation is determined in the gravitational field of an isolated, gravitationally bound N-body system. Based on the method of asymptotic perturbation theory, the various gravitational perturbations in phase also have been solved and the corresponding time delays are subsequently obtained. In addition to the conventional static fields, the solutions include the influence of motion, such as the velocity and acceleration. At the same time, we treat the system as not $N$ independent bodies but interaction-bound bodies. The gravitational interactions and tidal deformations have been taken into account for the realistic, nonrigid astronomic bodies. The condition of interaction bound is important in the strong gravitational fields, for example, tidal effects are significant and must be considered in the binary pulsar systems or in some GW sources.
Our solutions give a more precise description for physical N-body system and also are sufficient for the using in the modern space missions.
Applying them for the relativistic analysis in the TianQin mission, we specially focus on the gravitational influences of Earth-Moon system.  Based on the parameters of Keplerian orbit, we estimate the various terms in light propagation delays between spacecrafts (listed in TABLE.\ref{tab:1}), which may be used in a numerical model of TainQin. Eq.(\ref{phaseb}) or (\ref{piv2}) may be used to discuss the various-noise influences for TianQin sensitivity curve in the further works. From the relativistic analysis of TianQin, the Moon's gravitational effects on the onboard clocks and inter-spacecraft signal propagation are comparable with the Earth's, and the contributions of Earth-Moon-system gravitational delays are below the level of picometer at mHz regime that are negligible for GW detections.
Furthermore, we find that TianQin may provide some classical tests for general relativity.
For example, the uncertainty in the post-Newtonian parameter $\gamma$ may be tested at the level of testing some scalar-tensor theories of gravity and the Earth's gravitomagnetic effect on light propagation may be tested.

We have also computed the relativistic frequency shift between spacecraft's clocks due to the various motion and gravitational field terms. Parameterized estimates are listed in TABLE.\ref{tab:2}. We have found the surprising result that the general relativistic contribution in frequency shift is much smaller than $\emph{a}$ $\emph{priori}$ expectation.
These smaller contributions in frequency shift are mainly caused  by the TianQin configuration that makes several effects cancel with each other. At the same time, our results demonstrate that most of general relativistic effects are dependent on the orbital parameters. Thus, we can amplify or inhibit these effects through choosing an optimized orbit. These analytical formalism and parameterized estimates of the relativistic effects will provide a support for TianQin subsequent scientific mission.

\section{Acknowledgment}
The authors thank the anonymous referee for the useful comments on the improvement of the paper.
This work is supported by the National Natural Science Foundation of China(Grant Nos. 91636221 and 11805074).

\appendix

\section{eikonal equation}\label{appeikonal}
In order to solve the eikonal equation, the asymptotic perturbation method is used. We choose a small parameter $G$ (gravitational constant), and expand every function in the eikonal equation with the corresponding power series.
This method is similar to the approach developed by C. Le Poncin-Lafitte et al, which is initially based on the Synge world function \cite{sw} then on TTF \cite{ttf1}. The asymptotic power series can provide a safe way for selecting terms to keep at each order.
The metric tensor $g_{\mu\nu}$ is represented by a series in ascending powers of $G$:
\begin{equation}\label{ape1}
  g_{\mu\nu}(x,G)=\eta_{\mu\nu}+\sum^{\infty}_{n=1}G^{n}g^{(n)}_{\mu\nu}(x).
\end{equation}
Also, $g^{\mu\nu}$ can be given by a similar expansion with the relationships
\begin{equation}\label{ape2}
  g^{\mu\nu}_{(1)}=-\eta^{\mu\alpha}\eta^{\nu\beta}g^{(1)}_{\alpha\beta}
\end{equation}
and
\begin{equation}\label{ape3}
  g^{\mu\nu}_{(n)}=-\eta^{\mu\alpha}\eta^{\nu\beta}g^{(n)}_{\alpha\beta}-\sum^{n-1}_{p=1}\eta^{\mu\alpha}g^{(p)}_{\alpha\beta}g^{\beta\nu}_{(n-p)}.
\end{equation}
Further, the phase $\varphi$ is expressed as a similar expansion
\begin{equation}\label{ape4}
  \varphi(t,\textbf{x})=\varphi_{0}+\int{k_{\mu}dx^{\mu}}+\sum^{\infty}_{n=1}G^{n}\varphi^{(n)}(t,\textbf{x}),
\end{equation}
where $\varphi_{0}$ is a constant, and $\varphi^{(n)}(t,\textbf{x})$ is the phase perturbation of the $n$th order in $G$.

Let us define a light connecting $x_{A}=(ct_{A},\textbf{x}_{A})$ and $x_{B}=(ct_{B},\textbf{x}_{B})$ with point event $x=(|\textbf{x}-\textbf{x}_{A}|+\Delta,\textbf{x})$, where $\Delta$ is the gravitational delay, and $\textbf{x}$ is defined by parameter equation (\ref{paraeq}). Inserting Eqs.(\ref{ape2})-(\ref{ape4}) into the eikonal equation, we have the Hamilton-Jacobi-like equation
\begin{equation}\label{ape5}
  g^{\mu\nu}(x)\partial_{\mu}\varphi(x_{A},x)\partial_{\nu}\varphi(x_{A},x)=0.
\end{equation}
Using above equations, the perturbation terms $\varphi^{(n)}$ can be determined by a recursive procedure. Each term $\varphi^{(n)}$ can be given by the integral along a straight line between transmission and reception. It avoids the calculation of the gravitational perturbation of the geodesic joining the given points. In contrast, the integrals for $\varphi^{(n)}$ have to contain products of the first-order derivatives of lower-order terms $\varphi^{(n-p)}$, $n=1,...n-1$. It means that the calculations of integrals along the null geodesic are replaced by the calculations of the integrals of these derivatives. The method has been demonstrated by Poncin-Lafitte $et$ $al$ \cite{sw,ttf1}.
When we only consider the case of $n\leq2$, the solution of eikonal equation is expressed as \cite{ttf1,eik1}
\begin{equation}\label{ape6}
  \varphi^{(1)}=-\frac{R_{AB}}{2k_{0}}\int^{1}_{0}g^{\mu\nu}_{(1)}k_{\mu}k_{\nu}d\lambda
\end{equation}
and
\begin{eqnarray}\label{ape7}
  \varphi^{(2)}&&\!\!\!\!=-\frac{R_{AB}}{2k_{0}}\int^{1}_{0}(\eta^{\mu\nu}\partial_{\mu}\varphi^{(1)}\partial_{\nu}\varphi^{(1)}\nonumber\\
  &&\!\!\!\!+2g^{\mu\nu}_{(1)}k_{\mu}\partial_{\nu}\varphi^{(1)}+g^{\mu\nu}_{(2)}k_{\mu}k_{\nu})d\lambda.
\end{eqnarray}
All integrals are calculated along the straight line defned by Eq.(\ref{paraeq}).
Obviously, Eq.(\ref{ape6}) is the terms of the order of $G$ or $c^{-2}$, in which integral along straight is valid. Eq.(\ref{ape7}) is the order of $c^{-4}$. For the calculations of $\varphi^{(2)}$, the recursive procedure with terms $\partial_{\nu}\varphi^{(1)}$ avoids the integral along the perturbed paths.

\section{instantaneous coordinate distance}\label{apptime}
In the space missions, the recording time of signal is in one of satellites. Therefore, it is convenient to express the distance between spacecrafts at the time of reception $t_{B}$. We introduce an instantaneous coordinate distance $D_{AB}=|\textbf{D}_{AB}|=|\textbf{x}_{B}(t_{B})-\textbf{x}_{A}(t_{B})|$. By using the Taylor expansion, the coordinate distance $\textbf{R}_{AB}=\textbf{x}_{B}(t_{B})-\textbf{x}_{A}(t_{A})$ can be written as
\begin{equation}\label{app1}
\textbf{R}_{AB}
=\textbf{D}_{AB}+\textbf{v}_{A}(t_{B})T_{AB}-\frac{1}{2}\textbf{a}_{A}(t_{B})T^{2}_{AB}+O(c^{-3}),
\end{equation}
where $\textbf{v}_{A}$ and $\textbf{a}_{A}$ are the velocity and acceleration of $A$ at the coordinate time $t_{B}$, respectively. By an iterative process, $R_{AB}$ can be rewritten as
\begin{eqnarray}\label{app2}
  R_{AB}\!\!\!\!&&=D_{AB}+\frac{\textbf{D}_{AB}\cdot\textbf{v}_{A}}{c}+\frac{D_{AB}}{2c^{2}}\Big{[}\textbf{v}_{A}^{2}\nonumber\\
  &&-\textbf{D}_{AB}\cdot\textbf{a}_{A}+\frac{(\textbf{D}_{AB}\cdot\textbf{v}_{A})^{2}}{D^{2}_{AB}}\Big{]}+O(c^{-3}),
\end{eqnarray}
where all the quantities are measured at reception instant time $t_{B}$.
Then, we can obtain its derivative
\begin{equation}\label{app3}
  \frac{d R_{AB}}{cdt_{B}}=\frac{\textbf{n}_{AB}\cdot\textbf{v}_{AB}}{c}+\frac{1}{c^{2}}(\textbf{v}_{AB}\cdot\textbf{v}_{A}+\textbf{D}_{AB}\cdot\textbf{a}_{A})+O(c^{-3})
\end{equation}
with $\textbf{n}_{AB}=\textbf{D}_{AB}/D_{AB}$. Combining Eqs.(\ref{app1}) and (\ref{app2}), the unit vector $\textbf{n}_{AB}$ may be expressed in the terms of unit vector $\textbf{N}_{AB}$
\begin{eqnarray}\label{app4}
  \textbf{n}_{AB}&&\!\!\!\!\!\!=\textbf{N}_{AB}\Big{[}1+\frac{\textbf{N}_{AB}\cdot\textbf{v}_{A}}{c}\nonumber\\
 && \!\!\!\!\!\!+\frac{1}{2c^{2}}(3(\textbf{N}_{AB}\cdot\textbf{v}_{A})^{2}-\textbf{v}_{A}^{2}
  -\textbf{R}_{AB}\cdot\textbf{a}_{A})\Big{]}\\
   && \!\!\!\!\!\!-\frac{\textbf{v}_{A}}{c}\left (1+ \frac{\textbf{N}_{AB}\cdot\textbf{v}_{A}}{c}\right)+\frac{\textbf{a}_{A}}{2c^{2}}R_{AB}+O(c^{-3}).\nonumber
\end{eqnarray}
Using it, Eq.(\ref{app3}) can be rewritten as
\begin{eqnarray}\label{app5}
  \frac{d R_{AB}}{cdt_{B}}&&\!\!\!\!\!\!=\frac{\textbf{N}_{AB}\cdot\textbf{v}_{AB}}{c}
  +\frac{1}{c^{2}}[(\textbf{N}_{AB}\cdot\textbf{v}_{A})(\textbf{N}_{AB}\cdot\textbf{v}_{B})\nonumber\\
  &&\!\!\!\!\!\!-(\textbf{N}_{AB}\cdot\textbf{v}_{A})^{2}+\textbf{R}_{AB}\cdot\textbf{a}_{A}]+O(c^{-3}).
\end{eqnarray}
Neglecting the acceleration term, it recovers usual Doppler-effect form. And this method allows us to give the higher Doppler terms.

\section{Useful relationships for Keplerian orbit}\label{kepler}
We consider a Keplerian equation $r=a(1-e\cos u)$ in the orbital plane, where $a$ is the semimajor axis, $e$ is the eccentricity, and $u$ is the eccentric anomaly. In the orbital plane, the position vector is given by
\begin{equation}\label{k1}
  \textbf{r}=a(\cos u -e, \sqrt{1-e^{2}}\sin u).
\end{equation}
By this equation, its unit vector is given by
\begin{equation}\label{k2}
  \textbf{n}=\frac{\textbf{r}}{r}=(\frac{\cos u -e}{1-e\cos u}, \frac{\sqrt{1-e^{2}}\sin u}{1-e\cos u}).
\end{equation}
Considering the time derivative of the eccentric anomaly $\dot{u}=\sqrt{GMa}/ar$, the velocity vector is
\begin{equation}\label{k3}
  \textbf{v}=\frac{\sqrt{GMa}}{r}(-\sin u, \sqrt{1-e^{2}}\cos u).
\end{equation}
We consider two spacecrafts $A$ and $B$ with different eccentric anomalies $u_{A}$ and $u_{B}$. Using equations from (\ref{k1}) to (\ref{k3}), several relationships in the order of $e$ are
\begin{equation}\label{k4}
  \frac{\textbf{r}_{AB}\cdot\textbf{v}_{AB}}{r_{AB}}=-2\sqrt{\frac{GM}{a}}e\sin \mathcal{K}_{AB}\cos \mathcal{K}_{AB} \sin \mathcal{L}_{AB},
\end{equation}
\begin{equation}\label{k5}
  \frac{\textbf{r}_{AB}\cdot\textbf{r}_{A}}{r_{AB}}=-a(\sin \mathcal{K}_{AB}-e\sin \mathcal{L}_{AB}),
\end{equation}
\begin{equation}\label{k6}
   \frac{\textbf{r}_{AB}\cdot\textbf{v}_{A}}{r_{AB}}=\sqrt{\frac{GM}{a}}\cos \mathcal{K}_{AB}(1+e \cos u_{A}),
\end{equation}
with $r_{AB}=|\textbf{r}_{AB}|=|\textbf{r}_{B}-\textbf{r}_{A}|$ ,$\mathcal{K}_{AB}=(u_{B}-u_{A})/2$, and $\mathcal{L}_{AB}=(u_{B}+u_{A})/2$.


\begin{thebibliography}{}
\bibitem{pn1} S. Zschocke, Phys. Rev. D {\bf 92}, 063015 (2015).
\bibitem{pn2} S. Zschocke, Phys. Rev. D {\bf 93}, 103010 (2016).
\bibitem{pn3} S. Zschocke, Phys. Rev. D {\bf 94}, 124007 (2016).
\bibitem{pm1} S. M. Kopeikin, J. Math. Phys. (N.Y.) {\bf 38}, 2587 (1997).
\bibitem{pm2} S. M. Kopeikin, P. Korobkov, and A. Polnarev, Class, Quantum Grav. {\bf 23}, 4299 (2006).
\bibitem{pm3} C. Le Poncin-Lafitte and P. Teyssandier, Phys. Rev. D {\bf 77}, 044029 (2008).
\bibitem{sw} C. Le Poncin-Lafitte, B. Linet and P. Teyssandier, Class. Quantum Grav. {\bf 21}, 4463 (2004).
\bibitem{ttf1} P. Teyssandier and C. Le Poncin-Lafitte, Class. Quantum Grav. {\bf 25}, 145020 (2008).
\bibitem{ttf2} A. Hees, S. Bertone, and C. Le Poncin-Lafitte, Phys. Rev. D {\bf 89}, 064045 (2014).
\bibitem{3pm} B. Linet and P. Teyssandier, Class. Quantum Grav. {\bf 30}, 175008 (2013).
\bibitem{1} B. P. Abbott et al., Phys. Rev. Lett. {\bf 116} 061102 (2016).
\bibitem{2} B. P. Abbott et al., Phys. Rev. Lett. {\bf 116} 241103 (2016).
\bibitem{thr3} B. P. Abbott et al., Phys. Rev. Lett. {\bf 119} 141101 (2017).
\bibitem{p1} A. Nishizawa, Phys. Rev. D {\bf 97} 104037 (2018).
\bibitem{p2} S. Arai and A. Nishizawa, Phys. Rev. D {\bf 97} 104038 (2018).
\bibitem{3} B. P. Abbott et al., Phys. Rev. Lett. {\bf 119} 161101 (2017).
\bibitem{4} B. P. Abbott et al., Phys. Rev. Lett. {\bf 118} 221101 (2017).
\bibitem{5} P. Bender et al,. LISA (Laser Interferometer Space Antenna): An international project in the field of Fundamental Physics in Space (Max-Planck-Institute fur Quantenoptic, Garching bei Munchen, 1998).
\bibitem{sa} R. Hellings, SAGITTARIUS: an ESA M3 Proposal[J]. 1993.
\bibitem{om} T.R. Boehly et al., Optics Communication {\bf 133}, 495-506 (1997).
\bibitem{6} J. Luo et al., Class. Quantum Grav. {\bf 33}, 035010 (2015).
\bibitem{tdi0} N. J. Cornish, and R. W. Hellings, Class. Quantum Grav. {\bf 20 }, 4851 (2003).
\bibitem{tdi1} J. W. Armstrong, F. B. Estabrook, and M. Tinto, Phys. Rev. D {\bf 62}, 042002 (2000).
\bibitem{tdi2} S. V. Dhurandhar, K. R. Nayak, and J-Y. Vinet, Phys. Rev. D {\bf 65}, 102002 (2002).
\bibitem{tdi3} M. Tinto and O. Hartwig,  Phys. Rev. D {\bf 98}, 042003 (2018).
\bibitem{7} N. Ashby and B. Bertotti, Class. Quantum Grav. {\bf 27 },145013 (2010).
\bibitem{dsx1} T. Damour, M. Soffel and C. Xu, Phys. Rev. D {\bf 43} 3273 (1991).
\bibitem{dsx2} T. Damour, M. Soffel and C. Xu, Phys. Rev. D {\bf 45} 1017 (1992).
\bibitem{eik1} C.G. Qin, and C.G. Shao, Phys. Rev. D {\bf 96}, 024003 (2017).
\bibitem{8} B. Chauvineau et al., Phys. Rev. D {\bf 72} 122003 (2005).
\bibitem{gra} S.G. Turyshev, M.A. Sazhin, and V.T. Toth, Phys. Rev. D {\bf 89}, 105029 (2014).
\bibitem{aces}  S.G. Turyshev, N.Yu, and V.T. Toth, Phys. Rev. D {\bf 93}, 045027 (2016).
\bibitem{pnp} J. Sakstein, Phys. Rev. D {\bf 97} 064028 (2018).
\bibitem{fp1} R. Shaniv et al., Phys. Rev. Lett. {\bf 120} 103202 (2018).
\bibitem{fp2} C.G. Shao et al., Phys. Rev. D {\bf 97},  024019 (2018).
\bibitem{iau} M. Soffel, S. A. Klioner, G. Petit et al., Astrophys. J. {\bf126}, 2687 (2003).
\bibitem{space1} S.G. Turyshev, V.T. Toth, and M.A. Sazhin, Phys. Rev. D {\bf 87}, 024020 (2013).
\bibitem{eikonal1} V. Perlick, C. L$\ddot{\text{a}}$mmerzahl and A. Mac$\acute{\text{l}}$as, Phys. Rev. D {\bf 98}, 105014 (2018).
\bibitem{gwill} E. Poisson, and C.M. Will, \emph{Gravity: Newonian, Post-Newtonian, Relativistic} (Cambridge: Cambridge University Press, 2014), p. 119-p. 134.
\bibitem{egm8} N.K. Pavlis, S.A. Holmes, S.C. Kenyon, and J.K. Factor, J. Geophys. Res. {\bf 117} B04406 (2012), doi:10.1029/2011JB008916.
\bibitem{cassini} B. Bertotti, L. Iess, and P. Tortora, Nature  {\bf 425}, 374 (2003).
\bibitem{st1} T. Damour and K. Nordtvedt, Phys. Rev. Lett. {\bf 70}, 2217 (1993).
\bibitem{st2} J.E. Plowman and R.W. Hellings, Class. Quantum Grav. {\bf 23}, 309-318 (2006).
\bibitem{bc} S.G. Turyshev, M. Shao, A. Girerd, and B. Lane, Int. J. Mod. Phys. D {\bf 18}, 1025 (2009).
\bibitem{nob1} F. Riehle, Physics {\bf 5}, 126 (2012).
https://physics.aps.org/articles/pdf/10.1103/Physics.5.126
\bibitem{nob2} T.W. H$\ddot{\text{a}}$nsch, Rev. Mod. Phys. {\bf 78} 1297 (2006).
\bibitem{cqg2019} C.G. Qin, Y.J. Tan and C.G. Shao, Class. Quantum Grav. {\bf 36}, 055008 (2019).
\bibitem{c4} B. Linet, and P. Teyssandier, Phys. Rev. D {\bf 66}, 024045 (2002).
\end{thebibliography}
\end{document}